\documentclass[11pt]{article}
 \usepackage{amsfonts}
 \usepackage{amssymb}
 \def\bdt{\dot \beta}
 \def\adt{\dot \alpha}

 \newfont{\bbbold}{msbm10 scaled \magstep1}

 \def\bbC{\mbox{\bbbold C}}

 \def\bbH{\mbox{\bbbold H}}

 \def\bbK{\mbox{\bbbold K}}

 \def\bbO{\mbox{\bbbold O}}

 \def\bbR{\mbox{\bbbold R}}

 \def\cD{{\cal D}}
 
 \def\cF{{\cal F}}

 \newfont{\goth}{eufm10 scaled \magstep1}

 \def\a{\alpha}
 \def\b{\beta}
 \def\c{\gamma}\def\C{\Gamma}
 \def\d{\delta}
 \def\e{\epsilon}
 \def\vf{\varphi}
 \def\h{\eta}
 \def\k{\kappa}
 \def\l{\lambda}\def\L{\Lambda}
 \def\m{\mu}

 \def\s{\sigma}
 \def\t{\tau}
 \def\th{\theta}

 \def\del{\partial}
 \def\ua{\underline{\alpha}}
 \def\ub{\underline{\phantom{\alpha}}\!\!\!\beta}

 \def\una{\underline a}\def\unA{\underline A}
 \def\unb{\underline b}\def\unB{\underline B}
 \def\unC{\underline C}

 \def\unF{\underline F}
 
 \def\unM{\underline M}

 \def\unH{\underline{H}}\def\unG{\underline{G}}
 \def\unF{\underline{F}}

 \def\del{\partial}

 \def\be{\begin{equation}}\def\ee{\end{equation}}
 \def\bea{\begin{eqnarray}}\def\eea{\end{eqnarray}}
 \def\ba{\begin{array}}\def\ea{\end{array}}
 \let\la=\label
 \newcommand{\eq}[1]{(\ref{#1})}
 
 \def\det{{\rm det\,}}

\begin{document}

 \thispagestyle{empty}

 \hfill{KCL-TH-01-05}

 \hfill{hep-th/0103191}

 \hfill{\today}

 \vspace{20pt}

 \begin{center}
 {\Large{\bf Codimension zero superembeddings}}
 \vspace{30pt}

 {J.M. Drummond and P.S. Howe}
\vskip 1cm {Department of Mathematics} \vskip 1cm {King's College,
London} \vspace{15pt}

 \vspace{60pt}

 {\bf Abstract}

 \end{center}

Superembeddings which have bosonic codimension zero are studied in
3,4 and 6 dimensions. The worldvolume multiplets of these branes
are off-shell vector multiplets in these dimensions, and their self-interactions include a Born-Infeld term. It is shown
how they can be written in terms of standard vector multiplets in
flat superspace by working in the static gauge. The action formula
is used to determine both Green-Schwarz type actions and superfield
actions.

{\vfill\leftline{}\vfill \vskip  10pt

 \baselineskip=15pt \pagebreak \setcounter{page}{1}


 \section{Introduction}

 The superembedding formalism provides a powerful and systematic
 method for deriving the dynamics of super $p$-branes. A detailed review
 of the subject
 is given in \cite{d.s} where a comprehensive list of references
 can be found. There is a natural constraint on such an
 embedding, namely that the odd tangent space of the brane should
 be a subspace of the odd tangent space of the target superspace at
 all points on the brane. It was noted in \cite{hs} that this constraint
 is applicable to all branes, including those which have
 worldvolume multiplets containing gauge fields such as D-branes
 and the M5-brane, and that it
 determines a worldvolume multiplet which
 can be one of three distinct types.

 The three types of multiplet are: (a) on-shell multiplets, that is
 the basic embedding condition determines the dynamics, (b)
 off-shell multiplets where the embedding condition leads to a
 recognisable off-shell multiplet for which a superfield action can
 be written down, and (c) underconstrained multiplets. By
 underconstrained we mean that, although the worldvolume multiplet
 is off-shell, the resulting multiplet cannot be used to construct
 a Lagrangian. In the case where the target space has 32
 supersymmetries (and the worldvolume 16) the multiplets are mostly
 of type (a) but occasionally of type (c), the latter situation
 occurring for low (bosonic) codimension. For fewer supersymmetries
 the worldvolume multiplets are more often of type (b), but again
 type (c) multiplets arise for low codimension. For type (c),
 an additional constraint is required, and in all cases studied so
 far it turns out that it is sufficient to impose the so-called
 $\cF$-constraint, that is, one introduces an extra gauge field on
 the worldvolume and then constrains all of its components to
 vanish except for the purely bosonic ones. (For type (a) or (b)
 such gauge fields, if present, do not have to be introduced
 independently). If one considers
 systems of branes  the
 $\cF$-constraint can be derived from considerations of branes
 ending on other branes \cite{cs,csh,cshw}. In \cite{hkss} a detailed study was
 made of superembeddings with codimension one, where the
 worldvolume multiplet is an unconstrained scalar superfield.
 Except for the case of the membrane in $D=4$, which is type (b),
 these superembeddings are of type (c) and so it is necessary to
 impose the $\cF$-constraint. Alternatively, one can impose further
 constraints directly on the superembedding which are equivalent to
 the $\cF$-constraint, but it is simpler to impose the latter
 directly as it is unambiguous. An example of this is given by the
 5-brane in $D=7$ for which the equations were obtained by an
 additional geometrical constraint in \cite{acghn} and then by the
 $\cF$-constraint in \cite{hkss}.

 One can also derive actions starting from the superembedding
 formalism. Green-Schwarz type actions for branes can be obtained
 using the generalised action principle \cite{bsv} which was reinterptreted
 in \cite{hrs} as a constructive principle. For multiplets of type (a) one
 can also find superfield actions, the first model studied, the
 superparticle in $D=3$ being an example of this \cite{stv} (see also \cite{stvz} and \cite{vz} for other early papers). A third
 possibility, also requiring type (a) multiplets, is to construct
 actions in the static gauge. This has been done, for example, for the supermembrane in $N=1, D=4$  superspace indirectly in \cite{agit}, using partially broken supersymmetry systematically in \cite{ik} and more recently, starting from the superembedding approach in \cite{pst}.

 In this paper we consider superembeddings with bosonic codimension
 zero for spacetime dimensions three, four and six. The worldvolume
 multiplets are off-shell Maxwell multiplets in these dimensions
 and
 consist of a spinor field, a Maxwell field strength tensor and
 zero, one or three auxiliary scalars respectively
 combined together in a spinor superfield which can be identified with
 the transverse fermionic coordinate. The case of the
 D9-brane in ten dimensions has been studied previously in \cite{abkz};
 it differs from the others in the series in that the worldvolume
 multiplet is on-shell. A special feature of codimension zero is
 that, for these embeddings, the standard embedding condition
 imposes no constraint at all; one can always make a choice of the
 odd tangent bundle of the brane in such a way that it sits inside
 the odd tangent bundle of the target restricted to the brane.
 However, it still turns out that the $\cF$-constraint is
 sufficient to generate the required off-shell multiplets.

 It is quite simple to understand how this arises. Since the
 standard embedding condition imposes no constraints on the
 worldvolume multiplet, the latter must be an unconstrained spinor
 superfield corresponding to the transverse fermionic coordinate.
 One then introduces a new worldvolume Maxwell field with modified
 field strength $\cF$. In each of the three cases the standard
 constraint in flat superspace is that the odd-odd part of the
 field strength should vanish, so if we impose this (on $\cF$)
 there will be an off-shell Maxwell multiplet in addition to the
 spinor superfield. If we now require that the even-odd component of
 the field strength should vanish as well we eliminate one of the
 spinor fields, and thereby equate the (fermionic) Goldstone field
 of the
 superembedding with the field strength superfield of the Maxwell
 multiplet. Moreover, it is the unique covariant constraint which has this property.

 The notion of a brane in a flat superspace is directly related 
 to the notion of partial
 breaking of supersymmetry (specifically by one-half) \cite{hlp}. In a   superspace context this idea can be implemented using the group-theoretical method of non-linear realisations for supersymmetry \cite{va}. It is related to the superembedding formalism in that the former can be derived from the latter by working in a suitable gauge, and in some recent papers \cite{pst} membranes and the $N=2,D=2$ superparticle have been discussed from this point of view.
 An advantage of the
 superembedding formalism is that it can be applied to
 arbitrary target superspaces, although, as in the case of the
 $\k$-symmetric Green-Schwarz formalism \cite{bst}, the presence of branes
 may lead to constraints on the target superspace.

 The non-linear realisation method has
 been applied to many examples; see, for example \cite{agit,bg,rt, bik,dik,caip,p.w}. In the present
 context its relevance is that the worldvolume theory should then
 presumably be
 supersymmetric Born-Infeld theory in superspace. This has been
 studied in a number of papers, for example, \cite{cf, s.k,rtz,bik2}. In \cite{bs}, some partial results were given
 for higher-dimensional Yang-Mills theories, and a general analysis
 of the ten-dimensional superspace Bianchi identities which should
 be compatible with Born-Infeld theory has also appeared \cite{cnt}. Some
 of these results have included work on non-Abelian extensions of
 Born-Infeld and this is one of the main motivations for the
 current study. The hope is that by discussing the theory in a
 superembedding context one might be able to gain some insight into how to derive the non-Abelian
 generalisation which would be the right one for branes (although
 it is not known whether it would be unique). A discussion of this
 problem from the point of view of $\k$-symmetry has been given in
 \cite{bdr}. In fact, in the current paper, we shall not have much to say
 about the non-Abelian case. However, in order to be able
 to study this problem from the superembedding point of view, it is necessary to understand the abelian
 case first. Even here our results, obtained by going to the static
 gauge, are as yet incomplete. We are able to derive superspace
 Lagrangians for $N=1, D=3$ and $D=4$, but we have not yet made direct comparisons with the known
 results mentioned above. In order to do this, it is necessary to
 implemement the field redefinitions relating the different formalisms; work on this is in progress.
 The $D=6$ case is yet more complicated, and we also hope to study
 this problem in more detail in the near future.

 One aspect that we are able to comment on is the geometry induced
 on the worldvolume of the brane. For the $D=3$ case this is not
 terribly exciting, but for $D=4$ we find that there is an induced
 chiral supergeometry for which the Ogievetsky-Sokatchev potential
 $H^{\a\adt}$ \cite{os} (see also \cite{sg}), or rather its deviation from the flat case,
 is reminiscent of the supercurrent multiplet of the worldvolume gauge  supermultiplet. One
 might anticipate that there could be a harmonic superpace
 extension of this result in the $D=6$ case where the corresponding
 supergravity potential structure is also known \cite{e.s}.

 \section{Superembeddings}

 We consider a superembedding $f:M\rightarrow \unM$.
 Our index conventions are as follows; coordinate
 indices are taken from the middle of the alphabet with capitals
 for all, Latin for bosonic and Greek for fermionic, $M=(m,\m)$,
 tangent space indices are taken in a similar fashion
 from the beginning of the
 aplphabet so that $A=(a,\a)$. The distinguished tangent space
 bases are related to coordinate bases by means of the
 supervielbein, $E_M{}^A$, and its inverse $E_A{}^M$. Coordinates
 are denoted $z^M=(x^m,\th^{\m})$. We use exactly the same notation
 for the target space but with all of the indices underlined.
 Indices for the normal bundle are denoted by primes, so that
 $A'=(a',\a')$.

 The embedding matrix is the derivative of $f$ referred to the
 preferred tangent frames, thus

  \be
  E_A{}^{\unA}:= E_A{}^M\del_M z^{\unM} E_{\unM}{}^{\unA}
  \la{2.1}
  \ee

 The basic embedding condition is

  \be
  E_{\a}{}^{\una}=0
  \ee

To see the content of this constraint we can consider a linearised
embedding in a flat target space in the static gauge. This gauge is
specified by identifying the coordinates of the brane with a subset
of the coordinates of the worldvolume, so that

 \bea
 x^{\una}&=&\cases {x^a \cr x^{a'}(x,\th)} \\
 \th^{\ua}&=&\cases{\th^{\a}\cr \th^{\a'}(x,\th)}
 \la{2.2}
 \eea

Since

 \be
 E^{\una}=d x^{\una} -{i\over2}d\th^{\ua} (\C^{\una})_{\ua\ub}
 \th^{\ub}
 \la{2.3}
 \ee

in flat space, it is easy to see that, to first order in the
transverse fields, the embedding condition implies

 \be
 D_{\a} X^{a'}= i(\C^{a'})_{\a\b'} \th^{\b'}
 \la{2.4}
 \ee

where

 \be
 X^{a'}=x^{a'} + {i\over2} \th^{\a} (\C^{a'})_{\a\b'} \th^{\b'}
 \la{2.5}
 \ee

The worldvolume multiplet is therefore described by a set of scalar
superfields equal in number to the bosonic codimension and obeying
the constraint \eq{2.4}. Clearly, for codimension one, one has a
single otherwise unconstrained scalar superfield, which gives rise
to a type (c) multiplet in general. For codimension zero, there are
no transverse scalars, and so one has an unconstrained spinor
superfield $\th^{\a'}(x,\th)$.

In the non-linear theory it is useful to study this multiplet in a
covariant fashion using the geometrical quantities that are
available. In order to do this it is first of all necessary to
parametrise the odd-odd part of the superembedding matrix
$E_{\a}{}^{\ua}$. Preferred bases for the odd tangent bundle $\unF$
will be acted on by a group $\unG$, which is either the spin group
or a product of the spin group with an internal symmetry group.
Without loss of generality we can write

 \be
 E_{\a}{}^{\ua}= u_{\a}{}^{\ua} + h_{\a}{}^{\b'} u_{\b'}{}^{\ua}
 \la{2.6}
 \ee

where $u$ is an element of the group $\unG$ which will depend on
the brane coordinates in general. In other words, we split the odd
target space basis into two, with the same dimension, but allow
this splitting to depend on where we are on the brane. We can also
parametrise the even-even part of the embedding matrix in terms of
the Lorentz transformation corresponding to the spin transformation
in $u$; so we can choose

 \be
 E_a{}^{\una} = u_a{}^{\una}
 \la{2.7}
 \ee

At this stage there is still a local $\unG$ symmetry; that is, if
we transform $u\mapsto gu, g\in \unG$, the frame $E_{\a}$ will go
into itself up to linear transformation as long as we also
transform $h\mapsto h'$ where

 \be
 h'_{\a}{}^{\b'}=(-g_{\a}{}^{\c'} +g_{\a}{}^{\d}
 h_{\d}{}^{\c'})(g_{\c'}{}^{\b'} - g_{\c'}{}^{\e}
 h_{\e}{}^{\b'})^{-1}
 \la{2.8}
 \ee

In other words, $h$ transforms projectively under $\unG$. This
symmetry was discussed for the 3-brane in six dimensions in
\cite{hkss}; it can be used to choose different gauges for
$E_{\a}{}^{\ua}$.

The field $h$ was first introduced in \cite{hs} and plays a crucial
r\"{o}le whenever there are gauge fields on the brane. It is related in
a non-linear fashion to the field strength of the gauge field for
D-branes, for example. For scalar branes $h$ vanishes except if
there are auxiliary scalar fields.

\section{Static gauge}

It is not difficult to extend the linearised analysis presented
above to the non-linear case when the target space is flat. This
has been discussed in some examples in \cite{}, and shows how the
superembedding formalism is related to the non-linear realisation formalism \cite{}.

We can always choose coordinates, at least locally on the brane, in
which the odd frame has the form

 \be
 E_{\a}=A_{\a}{}^{\b}(D_{\b} + \psi_{\b}{}^b \del_b)
 \la{3.1}
 \ee

where, since we are discussing flat space, we no longer need  to
distinguish coordinate indices. We may take the even basis vectors
to be

 \be
 E_{a}= B_a{}^b \del_b
 \la{3.2}
 \ee

The matrices $A$ and $B$ are included for convenience in order to
facilitate comparison with the covariant approach.

The dual form bases are

 \bea
 E^{\a}&= &e^{\b} (A^{-1})_{\b}{}^{\a} \\
 E^a &=& (e^b - e^{\b} \psi_{\b}{}^b) (B^{-1})_b{}^a
 \la{3.3}
 \eea

where $e^a, e^{\a}$ denote the standard bases of flat superspace,

 \bea
 e^{\a}&=& d\th^{\a} \\
 e^{a} &=& d x^a -{i\over2} d\th^{\a} (\C^a)_{\a\b} \th^{\b}
 \la{3.4}
 \eea

If we then pull back the target space preferred frames and express
the result in terms of the frame $E^A$, the basic embedding
condition gives rise to a number of results. Firstly, we find that
the field $\psi_{\a}{}^a$ is given by

 \be
 \psi_{\a}{}^a={i\over2}
 D_{\a}\th'\C^b\th'(\d_b{}^a-{i\over2}\del_b \th'\C^a\th')^{-1}
 \la{3.5}
 \ee

This expresion essentially determines the induced geometry on the
brane. We also find the non-linear version of \eq{2.4}; it is

 \be
 \cD_{\a} X^{a'}=i(\C^{a'})_{\a\b'} \th^{\b'}
 \la{3.6}
 \ee

where

 \be
 \cD_{\a}:=D_{\a} + \psi_{\a}{}^a\del_a
 \la{3.7}
 \ee

with $X^{a'}$ given by \eq{2.5} as before. For the embedding matrix we
find

 \be
 E_{\a}{}^{\ub}\rightarrow \cases{ E_{\a}{}^{\b}=A_{\a}{}^{\b} \cr
 E_{\a}{}^{\b'}=A_{\a}{}^{\c}\cD_{\c}\th^{\b'}}
 \la{3.8}
 \ee

as well as

 \be
 E_a{}^{\unb}\rightarrow\cases{E_a{}^b=B_a{}^c
 (\d_c{}^b-{i\over2}\del_c\th'\C^b\th')\cr
 E_a{}^{b'}=B_a{}^c \del_c X^{b'}}
 \la{3.9}
 \ee

and

 \be
 E_a{}^{\ub}\rightarrow\cases{ E_a{}^{\b}=0\cr
 E_a{}^{\b'}=B_a{}^b\del_b\th'^{\b'} }
 \la{3.10}
 \ee

The matrices $A$ and $B$ can then be determined by comparing with
the covariant forms for the embedding matrix so that

 \bea
 u_{\a}{}^{\b} + h_{\a}{}^{\c'}u_{\c'}{}^{\b} &=& A_{\a}{}^{\b} \\
 u_{\a}{}^{\b'} + h_{\a}{}^{\c'} u_{\c'}{}^{\b'} &=&
 A_{\a}{}^{\c}\cD_{\c} \th^{\b'}
 \la{3.11}
 \eea

while

 \bea
 u_a{}^b &=& B_a{}^c (\d_c{}^b-{i\over2}\del_c\th'\C^b\th') \\
 u_a{}^{b'}&=& B_a{}^c \del_c X^{b'}
 \la{3.12}
 \eea

For the codimension zero case we have the same set of equations
with the difference that those equations involving $a'$ indices are
no longer present. Since the bosonic tangent spaces for brane and
target are the same, it is permissible to take $u_a{}^b=\d_a{}^b$,
and so we find an explicit expression for $B$ in terms of the field
$\th'(x,\th)$:

 \be
 B_a{}^b=(\d_a{}^b-{i\over2}\del_a\th'\C^b\th')^{-1}
 \ee

The corresponding spin group matrix is also trivial but $u$ with
odd indices need not be because of the presence of an internal symmetry group.

\section{Codimension zero}

It is well-known that the super Maxwell multiplets in $D=3,4,6$ and
$10$ dimensions are associated with the division algebras
$\bbK=\bbR,\bbC,\bbH$ and $\bbO$ respectively \cite{kt,j.e}. The codimension
zero embeddings in these dimensions can be understood in this
light, too, particularly for the cases we discuss in this paper,
namely $D=3,4,6$. In all four cases we have superembeddings of an
$N=1$ worldvolume superspace into an $N=2$ target superspace. The
spin group in these dimensions can be viewed as $SL(2,\bbK)$, the
internal symmetry group of the target space is $U(2,\bbK)$ and the
internal symmetry group of the worldvolume is $U(1,\bbK)$, although
the internal symmetry groups for $D=10$ do not seem to fit into
this pattern. Translated into more standard language, the internal
symmetry groups are $SO(2), U(2)$ and $USp(4)$ for $N=2$ in $D=3,4$
and 6 respectively, and $1,U(1)$ and $USp(2)=SU(2)$ for $N=1$. In
the following we shall use standard notation for each of the three
cases, so that spinors  are real with two-components in $D=3$,
complex with two-components in $D=4$, and pseudo-Majorana-Weyl with
eight components subject to a reality condition in $D=6$.

In view of the fact that we may choose the even-even part of the
embedding matrix to be the unit matrix for codimension zero, and
take the Lorentz group factor of the group element $u$ which occurs
in the embedding matrix to be $1$, the odd tangent space basis
vectors on the brane can be written

 \be
 E_{\a}=v_1{}^i E_{\a i} + h_{\a}{}^{\b} v_{2}{}^j E_{\b j}
 \la{4.1}
 \ee

where $i=1,2$ for $D=3,4$, and

 \be
 E_{\a i}=v_i{}^I E_{\a I} + h_{\a i}{}^{\b j'} v_{j'}{}^J E_{\b J}
 \la{4.2}
 \ee

where $i=1,2$, and $I=1,2,3,4$ for $D=6$. In both of these
formulae, the basis vectors on the right-hand side are standard
frames for the target space which we shall take to be flat for
simplicity throughout this paper. The matrix $v$, composed of
$(v_1{}^i,v_2{}^i)$ for $D=3,4$ and of $(v_i{}^I,v_{i'}{}^I)$ for
$D=6$ is an element of the target space internal symmetry group.
Strictly, in equation \eq{4.1}, the spinor indices should be
four-component with the spinors obeying a pseudo-Majorana-Weyl
condition in the $D=4$ case, but when we have imposed the $\cF$-constraint 
we shall see that it can be interpreted in terms of
two-component complex spinors.

The task now is to compute the dimension zero torsion on the brane,

 \be
 T_{\a\b}{}^c=E_{\a}{}^{\ua} E_{\b}{}^{\ub} T_{\ua\ub}{}^c
 \la{4.3}
 \ee

and then to introduce $\cF$ satisfying the modified Bianchi
identity

 \be
 d \cF = -\unH
 \la{4.4}
 \ee

where $\unH$ denotes the pull-back of a closed three-form on the
target superspace, $\unH=d\unB$. One then imposes the $\cF$-constraint

 \be
 \cF_{\a\b}=\cF_{\a b}=0
 \la{4.5}
 \ee

and analyses \eq{4.4} at dimension zero. This completes the
determination of the multiplet as a non-linear super-Maxwell
multiplet, and shows how the fields $h$ and $v$ in \eq{4.1} or
$\eq{4.2}$ are related to the dimension zero fields in the Maxwell
multiplets, i.e. the field strength tensors and auxiliary scalars.
To find the action it then only remains to specify the Wess-Zumino
$D+1$-form.

\subsection{D=3}

For dimension three we choose the gauge $v=1$ in \eq{4.1} so that

 \be
 E_{\a} = E_{\a 1} + h_{\a}{}^{\b} E_{\b 2}
 \la{4.6}
 \ee

with

 \be
 h_{\a\b}=\e_{\a\b} k + (\c^a)_{\a\b}h_a
 \la{4.7}
 \ee

The target space dimension zero torsion is

 \be
 T_{\a i\b j}{}^c= -i\d_{ij}(\c^c)_{\a\b}
 \la{4.7.1}
 \ee

using which we find the worldvolume dimension zero torsion is

 \be
 T_{\a \b}{}^a = -i(\c^b)_{\a\b} m_b{}^a
 \la{4.8}
 \ee

where

 \be
 m_{ab}=f\h_{ab} - 2h_a h_b -2 \e_{abc} k h^c
 \la{4.9}
 \ee

with

 \be
 f=1+k^2 + h^2
 \la{4.10}
 \ee

The closed target space three-form $H$ can be chosen to be such
that its only non-vanishing component (in flat space) is

 \be
 H_{\a i\b j c}= -i(\c_c)_{\a\b} (\t_1)_{ij}
 \la{4.11}
 \ee

where $\t_1$ is the first Pauli matrix. When the $\cF$-constraint
has been imposed the dimension zero component of the $\cF$ Bianchi
identity \eq{4.4} reads

 \be
 T_{\a\b}{}^b \cF_{bc}= E_{\a}{}^{\ua} E_{\b}{}^{\ub} H_{\ua\ub c}
 \la{4.12}
 \ee

where $\cF_{ab}=\e_{abc}\cF^c$. From this we find that $k=0$ so
that $f=1+h^2$, and also that

 \be
 \cF_a= {2h_a\over(1+h^2)}
 \la{4.13}
 \ee

As promised, therefore, the $h_{\a}{}^{\b}$ field in the embedding
matrix is non-linearly related to the field strength tensor of the
Maxwell multiplet.

\subsection{D=4}

For $D=4$ we can choose $\unH$ to be such that its only
non-vanishing component is

 \be
 H_{\a i\bdt c}^{\phantom{\a i} j}= -i(\t_1)_i{}^j(\s_c)_{\a\bdt}
 \la{4.14}
 \ee

where $\t_1$ again denotes the first Pauli matrix. If we use the
gauge in which $v=1$, it is straightforward to show that the
embedding is chiral in the sense that, in two-component notation,
there is no $h_{\a}{}^{\bdt}$. We simply have

 \be
 E_{\a}=E_{\a 1} + h_{\a}{}^{\b} E_{\b 2}
 \la{4.15}
 \ee

and similarly for the dotted odd basis vectors which are obtained
by complex conjugation. Using this in the $\cF$ identity one can
show that $h_{\a}{}^{\a}$ is subject to one real algebraic
constraint while $h_{(\a\b)}$ is proportional to $\cF_{\a\b}$, so
that the embedding describes the correct degrees of freedom of the
off-shell $N=1, D=4$ Maxwell supermultiplet. (Here $\cF_{\a\b}$
refers to the self-dual part of $\cF_{ab}$ in spinor notation.)
However, for computational purposes it is easier to switch to the
gauge where $h_{\a}{}^{\a}=0$, which requires the introduction of a
non-trivial $v$. If we do this and then impose the $\cF$-constraint
we find, after making a choice of $U(1)$ gauge ($U(1)$ being the
internal symmetry group of the worldvolume), that

 \be
 v=\exp ({iu\over2}\t_1)
 \la{4.16}
 \ee

as well as

 \be
 h_{\a\b}=A\cF_{\a\b}
 \la{4.17}
 \ee

where $A$ is a complex function which satisfies

 \be
 1+f^2 A\bar A= -\bar A
 \la{4.18}
 \ee

where $\cF_{\a}{}^{\c}\cF_{\c\b}:=\e_{\a\b}f^2$. In this gauge the
auxiliary field is $u$ while $A$ is determined in terms of
$\cF_{ab}$ from \eq{4.18}.

\subsection{D=6}

For $D=6$ it is again convenient to choose a gauge in which $v$ is non-trivial. Implementing the $\cF$-constraint, where

\be
H_{\a I\b J}= -i(\c)_{\a\b} H_{IJ}
\la{4.19}
\ee

and with

\be
H_{IJ}=\left(\ba{cc}
0& \e_{ij'}\\
\e_{i'j}& 0 \ea\right)
\la{4.20}
\ee

we find

\be
E_{\a i}= v_i{}^J E_{\a J} + h_{\a}{}^{\b}\d_i{}^{j'} v_{j'}{}^J E_{\b J}
\la{4.21}
\ee

where

\be
v = \exp \, i u_r
\left( 
\begin{array} {rr} 
0 & \t_r \\
\t_r & 0 
\end{array}
\right)
\ee

where $\t_r, r=1,2,3$ are the Pauli matrices. The three fields $u_r$ can be identified as the triplet of auxiliary fields while $h_{\a}{}^{\b}$ is a non-linear function of $\cF_{ab}$. 

\section{Actions}

In this section we discuss Green-Schwarz actions and superfield
actions for codimension zero branes. We concentrate on the $D=3$
case for simplicity adding some comments on the other cases at the
end.

We recall that the Wess-Zumino form $W_{D+1}$ is  a closed
$D+1$-form which can be written in the explicit form
$W_{D+1}=dZ_{D}$ where $Z_D$ is a potential $D$-form constructed
from given target space fields and $\cF$. On the brane it is exact
and so can be written $W_{D+1}=d K_D$. The Lagrangian form is
$L_D=K_D-Z_D$ and is closed by construction. The Green-Schwarz
action for the brane is

 \be
 S_{GS}=\int d^D x\, \e^{m_1\ldots m_D} L_{m_1\ldots m_D}(x,\th=0)
 \la{5.1}
 \ee

where the integration is taking over the bosonic worldvolume $M_o$.
A simple argument shows that this action is $\k$-symmetric and
reparametrisation invariant on $M_o$ \cite{hrs}. This procedure works for all
branes, provided that the worldvolume is not type (c),  with the
exception of those whose worldvolume multiplets contain self-dual
tensor fields for which other techniques are available \cite{pst2}. In
the case where the worldvolume multiplet is off-shell one can
construct superfield actions using two different approaches. The
first method involves imposing the embedding condition by a
Lagrange multiplier, as in the superparticle in $D=10$ \cite{gs} or the
heterotic string \cite{dhgs}, for example. Presumably this approach can be
generalised to codimension zero, but here we shall focus on the
second approach which can be used to derive an interacting superfield Lagrangian
for the worldvolume multiplet in the static gauge. Superfield
actions in the static gauge have been derived for membranes in \cite{agit,ik,pst}.

\subsection{D=3}

For $D=3$ the Wess-Zumino form for our choice of $H_3$ is

 \be
 W_4=\unG_4 + \unG_2 \cF
 \la{5.2}
 \ee

where $\unG_2$ is a closed target space two-form and $\unG_4$ satisfies

 \be
 d\unG_4=\unG_2 \unH_3
 \la{5.3}
 \ee

we therefore have

 \bea
 \unG_2&=& d\unC_1 \\
 \unG_4&=& d\unC_3 - \unC_1 \unH_3
 \la{5.3.1}
 \eea

where $\unC_1$ and $\unC_3$ are potential forms and
$\unH_3$ being the three-form appearing in the $\cF$ Bianchi identity.
In flat target space the non-vanishing components of the $\unG$-forms
are

 \bea
 G_{\a i\b j}&=& -i\e_{\a\b}\e_{ij} \\
 G_{\a i\b j cd}&=&-i(\c_{cd})_{\a\b} (\t_3)_{ij}
 \la{5.4}
 \eea

As noted above $W_4$ can be written as $dK_3$. It is
straightforward to check that the only non-vanishing component of
$K_3$ is the purely even one $K_{abc}=\e_{abc}K$, where

 \be
 K={(1-h^2)\over(1+h^2)}
 \la{5.5}
 \ee

Moreover, given the relation \eq{4.13} between $h_a$ and $\cF_a$, it is
easy to check that $K$ is the Born-Infeld Lagrangian

 \be
 K=\sqrt{(\h_{ab} + \cF_{ab})}
 \la{5.6}
 \ee

To compute the kinetic part of the action one then has to convert
to a coordinate basis using $E_m{}^a$ evaluated at $\th=0$. This
gives a factor of the determinant of this bosonic worldvolume
dreibein, which is the dreibein for the usual GS metric. The GS
action is then completed by the Wess-Zumino term. However, for
codimension zero it is possible to arrange for $E_m{}^{\a}$ to
vanish, so that one only needs to find $Z_{abc}$ in order to find
the Wess-Zumino term. The even-even part of
the worldvolume supervielbein is

 \be
 E_m{}^a=\d_m{}^b (B^{-1})_b{}^a
 \la{5.6.1}
 \ee

so that in any gauge the Green-Schwarz Lagrangian for the system is

 \be
 L_{GS} =\left(\det (B^{-1}) L\right)|
 \la{5.6.2}
 \ee

where $L_{abc}:=\e_{abc}L$, and where the bar denotes evaluation of a superfield  at
$\th=0$.

The superfield Lagrangian is  more difficult to compute. We use the method advocated in \cite{ggks} to relate $x$ space actions to closed $D$-forms in superspace and work in
the static gauge discussed in section 3 with the matrix $A$ set
equal to the unit matrix. The basis forms are then

 \bea
 E^a&=& (e^b - e^{\b} \psi_{\b}{}^b)(B^{-1})_b{}^a \\
 E^{\a}&=& e^{\a}
 \la{5.7}
 \eea

where $e^A=(e^a,e^{\a})$ are the standard basis forms for $N=1,D=3$
flat superspace. The idea is now to identify a component of the
three-form $L_3$ as the superspace Lagrangian and then to show that
the resulting $x$-space action is the same as the GS action. This
is most easily done by working in the flat basis $e^A$, so we shall
write the components of $L_3$ in this basis as $\ell_{ABC}$,
whereas its components in the $E^A$ basis will be denoted by
$L_{ABC}$.

Now, if the Lagrangian three-form is changed by the addition of an
expression of the form $d X_2$, where $X_2$ is some two-form, the
action will be unaltered, so we can make use of this freedom to
change $L_3$ such

 \be
 \ell_{\a\b\c}=0
 \la{5.8}
 \ee

(Note that this is not the case for the original $L_3$.) This is
always possible by a suitable choice of $X_{\a b}$ since
$(dX)_{\a\b\c}$ includes a term of the form $(\c^d)_{(\a\b}
X_{\c)d}$; in fact, $\ell_{\a\b\c}$ can be made equal to zero by
such a transformation where $X_{\a b}$ is gamma-traceless. Since
the new $L$ is still closed it is straightforward to check that
this implies, for an appropriate choice of $X_{ab}$,

 \be
 \ell_{\a\b c}= (\c_c)_{\a\b}L_o
 \la{5.9}
 \ee

It is $L_o$ which is to be identified with the superspace
Lagrangian. To prove that this is correct we first note that the
scalar part is unaffected by the change of $L_3$ by $dX_2$ due to
the gamma-tracelessness of $X_{\a b}$ (It is also assumed that
$X_{\a\b}=0$.) Secondly, it is easy to show that the top component
of $L$, i.e. $\ell_{abc}:=\e_{abc} \ell$, is the top component of
the scalar superfield $L_o$ and is thus the $x$-space Lagrangian
for the superfield action

 \be
 S_{SF}=\int d^3x\,d^2\th\,L_o,
 \la{5.10}
 \ee

i.e. $\ell= D_{\a} D^{\a} L_o$. Now, under the change
$L_3\rightarrow L_3+ dX_2$, $\ell_{abc}$ does change, but by $\del_{[a}X_{bc]}$
and this is irrelevant in the ($x$-space) action.
On the other hand it is easy to see that

 \be
 L_{abc}= B_a{}^d B_b{}^e B_c{}^f \ell_{def}
 \la{5.11}
 \ee

from which we find that $\ell=\det(B^{-1})L=L_{GS}$  when evaluated at $\th=0$ as claimed.

To compute it explicitly it is necessary to make gauge choices for
the $\unC$ fields. Normally one would do this treating $\th^{\a 1}$
and $\th^{\a 2}$ on an equal footing. However, in the present
context we do not wish to have explicit worldvolume $x's$ or
$\th's$ appearing in the Lagrangian, so it is better to choose a
gauge in which the potentials only depend on $\th^2$, which becomes
the brane superfield, and which we shall denote by $\L(x,\th)$.

For the three-form potential $\unC_3$ we can choose a gauge in which
the non-vanishing components are

 \bea
 C_{\a 1\b 1c}&=& (\c_c)_{\a\b} (\th^2)^2\\
 C_{\a 2\b 2c}&=& -(\c_c)_{\a\b} (\th^2)^2\\
 C_{\a 2 bc}&=&i(\c_{bc})_{\a\b} \th^{\b 2}\\
 C_{abc}&=& \e_{abc}
 \la{5.12}
 \eea

where $\th^{\a 2}\th^{\b 2}:=\e^{\a\b}(\th^2)^2$. For $\unC_1$ we have

 \be
 C_{\a 1}=i\th_{\a}^2
 \la{5.13}
 \ee

Given these choices it is straightforward to find the Lagrangian
three-form and compute its components with respect to the flat basis. 
To find the superfield Lagrangian we
need

 \be
 L_o \propto (\c^c)^{\a\b}\ell_{\a\b c}
 \la{5.15}
 \ee

A shortish computation yields

 \be
 L_o=\L^2 \left(1-{1\over2}D_{\a}\L_{\b} D^{\a}\L^{\b}\right)^{-1}
 \ee

This final Lagrangian is in agreement with that derived in \cite{ik} using partially broken supersymmetry (after a field redefinition). 

\subsection{D=4,6}

For $D=4,6$ deriving the Green-Schwarz actions is a similar
procedure. In $D=4$ the Wess-Zumino form is

 \be
 W_5=\unG_5 + \unG_3 \cF
 \la{5.16}
 \ee

where $d\unG_5=\unG_3 \unH_3$. The non-vanishing components of the $\unG$'s are

 \bea
 G_{\a i\bdt c}^{\phantom{\a i} j}&=& -i(\s_c)_{\a\bdt}
 (\t_2)_i{}^j\\
 G_{\a i\bdt cde}^{\phantom{\a i} j}&=&-2(\s_{cde})_{\a\bdt}
 (\t_3)_i{}^j
 \la{5.17}
 \eea

Using the results of subsection 4.2 it is relatively
straightforward to show that $K_{abcd}:=\e_{abcd}K$ is the only
non-vanishing component of $K_4$ and that it has the explicit form

 \be
 K=\cos u\sqrt{-\det(\d_a{}^b + \cF_a{}^b)}
 \la{5.17.1}
 \ee

so that, on eliminating $u$, we recover the standard Born-Infeld form.

For $D=6$ we have

 \be
 W_7=\unG_7 + \unG_5\cF + {1\over2}\unG_3 \cF^2
 \la{5.18}
 \ee

where

 \bea
 d\unG_7&=& \unG_5 \unH_3\\
 d\unG_5&=& \unG_3 \unH_3
 \la{5.19}
 \eea

The non-vanishing components of the target space forms are

\bea
G_{\a I \b J c}& = &(\c_{c} ) _{\a \b} G^{3}_{I J}\\
G_{\a I \b J cde}& = &(\c_{cde} ) _{\a \b} G^{5}_{I J}\\
G_{\a I \b J cdefg}& = &\e_{cdefgh}(\c^{h} ) _{\a \b} G^{7}_{I J}
\eea

where the matrices $G^3$ , $G^5$ , $G^7$ are

\be G^{3}_{I J} =
i \left( 
\begin{array} {rr} 
\e_{i j} & 0\\
0 & -\e_{i^{\prime} j^{\prime}} 
\end{array}
\right)
\ee
\be G^{5}_{I J} =
i \left( 
\begin{array} {rr} 
0 & \e_{i j^{\prime}}\\
-\e_{i^{\prime} j} & 0
\end{array}
\right)
\ee
\be G^{7}_{I J} =
i \left( 
\begin{array} {rr} 
\e_{i j} & 0\\
0 & -\e_{i^{\prime} j^{\prime}} 
\end{array}
\right)
\ee

The $USp(4)$ invariant metric is 

\be \eta_{I J} =
\left( 
\begin{array} {rr} 
\e_{i j} & 0\\
0 & \e_{i^{\prime} j^{\prime}} 
\end{array}
\right)
\ee

In this case one can again verify $W_7=d K_6$, and that the only non-vanishing compoent of $K_6$ is the purely even one. We anticipate that we should again find the Born-Infeld Lagrangian dressed by the auxiliary scalar fields.

As in the $D=3$  case we can construct superfield actions 
although they are no longer full superspace integrals. It is easy
to see, for $D=4,6$, that $L_D$ does not contain a scalar component
of the right dimension to be a candidate Lagrangian for such an
action.

The $D=4$ case was studied in detail for a general theory in \cite{ggks}. In
flat superspace, for example, given a closed four-form $L_4$
satisfying the constraints

 \bea
 L_{\a\b\c\d}&=&0\\
 L_{\a\b\c d}&=&0
 \la{5.19.2}
 \eea

one can find a chiral Lagrangian in the component $L_{\adt\bdt cd}$
which includes a term of the form $(\s_{cd})_{\adt\bdt}L_o$. The
real part of the top component of this form is the purely even part
of $L_4$, i.e. $L$, where $L_{abcd}=\e_{abcd}L$, 
so that $L=D^2 L_o + \bar D^2 \bar L_o$. Again one is
allowed to change $L_4$ by $dX_3$ for some three-form $X_3$. To
apply this to the brane case, we can use a similar argument to the
the $D=3$ case to show that the chiral action constructed from
$L_4=K_4-Z_4$ in the flat basis $e^A$ reproduces the Green-Schwarz
action.

In the $D=6$ case, the superfield action we are able to construct
relatively easily is an example of a superaction \cite{hst}, discussed
for $D=6$ supersymmetric Yang-Mills theory in \cite{hs'}. This time,
in flat space, if all the lower components in $L_6$ vanish one can
show that

 \be
 L_{abcd \a i\b j}=(\c_{[abc})_{\a\b} L_{d] ij}
 \la{5.20}
 \ee

The field $L_{a ij}$ (symmetric on $ij$) is the candidate
Lagrangian; it is to be integrated with respect to the ``measure''
$d^6 x\, D_{\a}^iD_{\b}^j (\c^a)^{\a\b}$. The
resulting action is supersymmetric due to the fact that $dL_6=0$. For the brane we anticipate that the argument given
above for the $D=3$ and $D=4$ cases will work here as well,
although the full details remain to be worked out.

\section{Induced geometry}

In this section we discuss some aspects of the supergeometry
induced on the brane for codimension zero embeddings. In the case
of $D=3$, the choice of bases we made earlier led to a non-standard
form for the worldvolume dimension zero torsion. However, we can
easily bring it to standard form by a change of even basis using
the matrix $m_a{}^b$. One can then modify $E_a$ and fix the
$SL(2,\bbR)$ connection on the brane to obtain the standard form of
$D=3$ supergeometry. Essentially the only constraints one imposes
in $D=3$ are conventional ones. In the embedding framework, the
geometry is therefore completely standard and the induced
supergravity potential is the field $\psi_{\a}{}^b$ which is given
explicitly in terms of the transverse fermion field $\L_{\a}$.

For $D=4$ the situation is more interesting because, as we have
seen, the $\cF$-constraint enforces chirality. In two-component
notation the even target space basis forms are

 \be
 E^a=dx^a+{i\over2}d\th^{\a i}(\s^a)_{\a\adt}\bar\th^{\adt}_i +
 {i\over2}d\bar\th^{\adt}_i(\s^a)_{\a\adt}\th^{\a i}
 \ee

Pulling this back to the worldvolume in the gauge where $v=1$ and
exploiting chirality we find

 \be
 \cD_{\a} \bar\L^{\bdt}=0
 \la{5.22}
 \ee

where

 \be
 \cD_{\a}:=D_{\a} + \psi_{\a}{}^a\del_a
 \la{5.22.1}
 \ee

as before, with

 \be
 \psi_{\a}{}^a=-{i\over2}(D_{\a}\L\s^b\bar\L+D_{\a}\bar\L\s^b\L)
 (\d_b{}^a+ {i\over2}\del_b\L\s^s\bar\L +
 {i\over2}\del_b\bar\L\s^a\L)^{-1}
 \la{5.23}
 \ee

From these equations we see that the chirality constraint on $\L$
given in \eq{5.22} is highly non-linear. However, after a little
algebra (and using the fact that $\L$ is chiral) we find that
$\psi_{\a}{}^a$ can be rewritten as

 \be
 \psi_{\a}{}^a=iD_{\a} J^b (\d_b{}^a -i\del_b J^a)^{-1}
 \la{5.24}
 \ee

where

 \be
 J^a:=-{1\over2}\L\s^a\bar\L
 \la{5.24.1}
 \ee

From this, we find that the odd basis vector on the worldvolume is

 \be
 E_{\a}=D_{\a}+ iD_{\a} J^b (\d_b{}^a -i\del_b
 J^a)^{-1}\del_a
 \la{5.25}
 \ee

This is reminiscent of the form of $E_{\a}$ in Ogievetsky-Sokatchev
supergravity obtained by transforming from special chiral
coordinates to standard coordinates. To see this we recall that 
we can complexify an $N=1,D=4$ superspace $M$ with a chiral
structure to $M_L$, say, and then use Frobenius' theorem to write
the continuation of the dotted basis as

 \be
 E_{\adt}=-{\del\over\del\vf_L^{\adt}}
 \la{5.26}
 \ee

in adapted coordinates $(x_L,\th_L,\vf_L)$. These coordinates are
related to the analytic continuation of the coordinates of $M$ by

 \bea
 x_{L}^a&=&x^a + iH^a(x,\th,\bar\th) \\
 \th_L^{\a}&=&\th^{\a} \\
 \vf_L^{\adt}&=&\bar\th^{\adt}
 \la{5.27}
 \eea

where $H^a$ is the Ogievetsky-Sokatchev potential, and $\bar\th$
becomes the complex conjugate of $\th$ when we return to the real
superspace. When we express $E_{\adt}$ in terms of
$(x,\th,\bar\th)$ and return to real superspace we find (taking the
complex conjugate)

 \be
 E_{\a}={\del\over\del\th^{\a}}+i\del_{\a}H^b(\d_b{}^a-i\del_b
 H^a)^{-1}
 \la{5.28}
 \ee

This is the same as \eq{5.25} provided that we set

 \be
 H^{\a\adt}=-{1\over2}\th^{\a}\bar\th^{\adt}+J^{\a\adt}
 \la{5.29}
 \ee

Note that the first term in this expression is just the potential
for flat superspace. In other words the induced Born-Infeld
geometry for $D=4$ is an Ogievetsksy-Sokatchev geometry with
potential equal to the flat potential plus a term which looks to be
the supercurrent for the Maxwell multiplet.

\section{Conclusions}

In this paper we have described some aspects of superembeddings
with bosonic codimension zero in $D=3,4,6$. In particular, we have
discussed how the worldvolume super-Maxwell multiplets arise when
the $\cF$-constraint is imposed and we have seen how one can
construct Green-Schwarz type actions and superfield actions in the
static gauge. Further calculations will be necessary in order to
have a completely satisfactory description of these theories as
superembeddings, particularly from the point of view of comparing
the results that can be derived from this formalism with
superspace Born-Infeld theory.

One task that we have not carried out in this paper is to find the
worldvolume superspace field strength in the static gauge. If we
write $\cF=F-\unB_2$, then $F$ satisfies a normal Bianchi identity
$dF=0$. Its components can be computed quite straightforwardly
knowing $\unB_2$ and $\cF$. Again, it is preferable to choose a gauge
for $\unB_2$ on the target space such that its components depend only
on $\th^{2}$ and not on $x$ or $\th^1$. This is always possible,
although the resulting $F$ is complicated. Moreover, the field strength superfield $\L$ is not the standard one that one would use in flat superspace. For $D=4$, for example, the field $\L$ is covariantly chiral, whereas the usual field strength  is ordinarily chiral, as it is in \cite{cf}. However, it is not difficult to construct a chiral field $\l=\L+\dots..$ and this will be discussed in more detail elsewhere. (The relation between the chiral and covariantly chiral spinor superfields has been discussed in the context of partially broken supersymmetry in \cite{bg}.)

One generalisation that can be made concerns the target space
geometry. In this paper we have made the simplest choices possible
for the various forms that are needed, i.e. the $\unG's$ and $\unH's$,
but it is possible to treat these in a more systematic fashion. The
sets of forms that arise in this way presumably reflect the
supergravity theories which can provide consistent backgrounds for
these objects.

Finally, we have noted that the induced geometry for $D=4$ turns
out to have a simple interpretation in terms of the
Ogievetsky-Sokatchev potential. As we remarked earlier, it would be
interesting to see if this could be generalised to the $D=6$ case
where one might suspect that harmonic superspace methods could come
into play.

\section*{Acknowledgements}

This work was supported in part by PPARC through SPG grant 68.

\section*{Note added}

In a recent preprint some similar results have been derived for the D=4 case \cite{bppst}.


\begin{thebibliography}{99}


 
\bibitem{d.s}
D. Sorokin, 
{\sl Superbranes and superembeddings} Phys. Report {\bf 239} (2000) 1. 

\bibitem{hs}
P.S. Howe and E. Sezgin,
{\sl Superbranes} Phys. Lett. {\bf B390} (1997) 133-142, hep-th/9607227.

\bibitem{cs}
C.S. Chu and E. Sezgin,
{\sl M5 brane from the open supermembrane} JHEP {\bf 12} (1997) 001, hep-th/9710223.
 
\bibitem{csh}
C.S. Chu, P.S. Howe and E. Sezgin,
{\sl Strings and D-branes with boundaries} Phys. Lett. {\bf B428} (1998) 59-67, hep-th/9801202.

\bibitem{cshw}
C.S. Chu, P.S. Howe, E. Sezgin and P.C. West
{\sl Open superbranes} Phys. Lett. {\bf 429} (1998) 273-280, hep-th/9803041. 

\bibitem{hkss}
P.S. Howe, A. Kaya, E. Sezgin and P. Sundell,
{\sl Codimension one branes} Nucl. Phys. {\bf B587} (2000) 481-513, hep-th/0001169.

\bibitem{acghn}
T. Adawi, M. Cederwall, U. Gran, M. Holm and B.E.W. Nilsson,
{\sl Superembeddings, non-linear supersymmetry and 5-branes} Int. J. Mod. Phys. {\bf A13} (1998) 007, hep-th/9711203.

\bibitem{bsv}
I. Bandos, D. Sorokin and D.V. Volkov,
{\sl On the generalised action principle for superstrings and supermembranes} Phys. Lett. {\bf B352} (1995) 269.

\bibitem{hrs}
P.S. Howe, O. Raetzel and E. Sezgin
{\sl On brane actions and superembeddings} JHEP {\bf 9808} (1998) 011, hep-th/9804051. 

\bibitem{stv}
D. Sorokin, V. Tkach and D.V. Volkov,
{\sl Superparticles, twistors and Siegel symmetry} Mod. Phys. Lett. {\bf A4} (1989) 901. 

\bibitem{stvz}
D. Sorokin, V. Tkach, D.V. Volkov and A. Zheltukhin
{\sl From the superparticle Siegel symmetry to the spinning particle proper time supersymmetry}, Phys. Lett. {\bf B216} (1989) 302.

\bibitem{vz}
D.V. Volkov and A. Zheltukhin
{\sl Extension of the Penrose representation and its use to describe supersymmetric models} Sov. Phys. JETP Lett. {\bf 48} (1988) 63-66.

\bibitem{agit}
A. Achucarro, J. Gauntlett, K. Itoh and P. Townsend, {\sl Worldvolume supersymmetry from spacetime supersymmetry of the four-dimensional supermembrane}, Nucl. Phys. {\bf B314} (1989) 129.

\bibitem{ik}
E. Ivanov and S. Krivonos, {\sl N=1, D=4 supermembrane in the coset approach}, Phys. Lett. {\bf B453} (1999) 237-244, hep-th/9901003.

\bibitem{pst}
P. Pasti, D. Sorokin and M. Tonin,
{\sl Supersymmetry, partial supersymmetry breaking and branes}, hep-th/0007048; {\sl Geometrical aspects of superbrane dynamics}, hep-th/0011020.

\bibitem{abkz}
V. Akulov, I. Bandos, W. Kummer and V. Zima
{\sl D=10 Dirichlet super 9-brane}, hep-th/9802032. 

\bibitem{hlp}
J. Hughes, J. Liu and J. Polchinski, {\sl Supermembranes}
Phys. Lett. {\bf B180} (1986) 370; J. Hughes and J. Polchinski, {\sl Partially broken global supersymmetry and the superstring} Nucl. Phys. {\bf B278} (1986) 147.

\bibitem{va}
D.V. Volkov and V.P. Akulov, JETP Lett. {\bf 16} (1972) 438; Phys. Lett. {\bf B46} (1973) 109.

\bibitem{bg}
J. Bagger and A. Galperin, {\sl A new Goldstone multiplet for partially broken supersymmetry} Phys. Rev. {\bf D55} (1997) 1091-1098, hep-th/9608177; {\sl The tensor Goldstone multiplet for partially broken supersymmetry} Phys. Lett. {\bf B412} (1997) 296-300, hep-th/9707061.

\bibitem{rt}
M. Rocek and A. Tseytlin, {\sl Partial breaking of global D=4 supersymmetry, constrained superfields and 3-brane actions}, hep-th/9811232.

\bibitem{bik}
S. Bellucci, E. Ivanov and S. Krivonos, {\sl Partial breaking of N=1, D=10 supersymmetry} Phys. Lett. {\bf B460} (1999) 348, hep-th/9811244; {\sl Partial braeking of N=4 to N=2: hypermultiplet as a Goldstone superfield} Fortsch. Phys. {\bf 48} (2000) 19, hep-th/9809190; {\sl Superworldvolume dynamics of branes from non-linear realisations}, Phs. Lett. {\bf B482} (2000) 233, hep-th/0003273.

\bibitem{dik}
F. Delduc, E. Ivanov and S. Krivonos, {1/4 partial breaking of global supersymmetry and new superparticle actions}, Nucl. Phys. {\bf B576} (2000) 196-218, hep-th/9912222.

\bibitem{caip}
C. Chryssomalakos, J.A. de Azcarraga, J.M. Izquierdo and J.C. Perez Bueno, {\sl The geometry of branes and extended superspaces},  Nucl.Phys. {\bf B567} (2000) 293-330, hep-th/9904137. 

\bibitem{p.w}
P. West, {\sl Automorphisms, non-linear realisations and branes}, hep-th/0001216.

\bibitem{bst}
E. Bergshoeff, E. Sezgin and P. Townsend
{\sl Supermembranes and eleven-dimensional supergravity} Phys. Lett {\bf B189} (1987) 75-78. 

\bibitem{cf}
S. Cecotti and S. Ferrara
{\sl Supersymmetric Born-Infeld Lagrangians} Phys. Lett. {\bf B187} (1987) 335.

\bibitem{s.k}
S. Ketov
{\sl A manifestly N=2 supersymmteric Born-Infeld action}, hep-th/9809121: {\sl N=1 and N=2 Supersymmetric non-Abelian Born-Ineld actions from superspace}, hep-th/0005625.

\bibitem{rtz}
A. Refolli, N. Terzi and D. Zanon
{\sl Non-Abelian N=2 supersymmetric Born-Infeld action} Phys. Lett. {\bf B486} (2000) 337, hep-th/0006067. 

\bibitem{bik2}
S. Bellucci, E. Ivanov and S. Krivonos, 
{\sl N=2 and N=4 supersymmetric Born-Infeld theories from non-linear realisations}, Phys. Lett. {\sl B502} (2001) 279-290, hep-th/0012236;
{\sl Towards the complete N=2 superfield Born-Infeld action with partially broken N=4 supersymmetry}, hep-th/0101195. 

\bibitem{bs}
E. Bergshoeff and E. Sezgin,
{\sl Higher derivative super Yang-Mills theories}, Phys. Lett. {\bf B185} (1987) 371-376.

\bibitem{cnt}
M. Cederwall, B. Nilsson and D. Tsimpis, {\sl The structure of maximally supersymmetric Yang-Mills theory: constraining higher-order corrections}, hep-th/0102009; {\sl D=10 super Yang-Mills theory at order $\a'^2$}, hep-th/0104236.

\bibitem{bdr}
E. Bergshoeff, M. de Roo and A. Sevrin
{\sl Non-Abelian Born-Infeld and kappa-symmetry}, hep-th/0011018.


\bibitem{os}
V. Ogievetsky and E. Sokatchev,
{\sl The axial superfield and the supergravity group} Yad. Fiz. {\bf 28} (1978) 1631-1639; {\sl Equation of motion for the axial gravitational superfield}, Sov. J. Nucl. Phys.  {\bf 32} (1980) 447. 

\bibitem{sg}
W. Siegel and S.J. Gates
{\sl Superfield supergravity} Nucl. Phys. {\bf B147} (1979) 77-104.



\bibitem{e.s}
E. Sokatchev,
{\sl Off-shell six-dimensional supergravity in harmonic superspace} Class. Quant. Grav. {\bf 5} (1988) 1459-1471.


\bibitem{kt}
T. Kugo and P.K. Townsend,
{\sl Supersymmetry and the division algebras}, Nucl. Phys. {\bf B221} (1983) 357. 


\bibitem{j.e}
J. Evans,
{\sl Supersymmetric Yang-Mills theories and the division algebras} Nucl. Phys. {\bf B298} (1988) 92.

\bibitem{pst2}
P. Pasti, D. Sorokin and M. Tonin
{\sl Covariant action for the superfivebrane of M-theory} Phys. Lett. {\bf B398} (1997) 41-46, hep-th/9701037.
 
\bibitem{gs}
A. Galperin and E. Sokatchev
{\sl A twistor like D=10 superparticle action with manifest N=8 worldline supersymmetry} Phys. Rev. {\bf D46} (1992) 714-725, hep-th/9203051.

\bibitem{dhgs}
F. Delduc, A. Galperin, P.S. Howe and E. Sokatchev
{\sl A twistor formulation of the heterotic D=10 superstring with manifest (8,0) worldsheet supersymmetry} Phys. Rev {\bf D47} (1993) 578-593, hep-th/9207050..

\bibitem{ggks}
S.J. Gates, M. Grisaru, M. Knutt-Wehlau and W. Siegel,
{\sl Component actions from curved superspace: normal coordinates and ectoplasm} Phys.Lett. {\bf B421} (1998) 203-210, hep-th/9711151.
 
\bibitem{hst}
P.S. Howe, K.S. Stelle and P.K. Townsend
{\sl Superactions} Nucl. Phys. {\bf B191} (1981) 445-464.

\bibitem{hs'}
P.S. Howe and K.S. Stelle, 
{\sl The ultra-violet properties of supersymmetric field theories}, Int. J. Mod. Phys. {\bf A4} (1989) 1871-1913.

\bibitem{bppst}
I. Bandos, P. Pasti, A. Pokotilov, D. Sorokin and M. Tonin, {\sl The space-filling 3-brane in N=2, D=4 superspace}.

 
\end{thebibliography}
 \end{document}